\documentclass{aa}  

\usepackage{graphicx}
\usepackage{txfonts}
\usepackage[pdfencoding=auto,psdextra]{hyperref}
\hypersetup{
    colorlinks=true,
    linkcolor=blue,
    urlcolor=blue,
    citecolor=blue
}

\usepackage{lscape}             
\usepackage{placeins}          
                            
\begin{document}

   \title{Joint probabilistic inference of galaxy redshifts and rest-frame spectra from photometric fluxes with latent diffusion}



   \author{Han-Yue Guo\inst{1,2}\fnmsep\thanks{email: hguo@pic.es}
        \and Martin Eriksen\inst{1,2}
        }

   \institute{Institut de Física d’Altes Energies (IFAE), The Barcelona Institute of Science and Technology, Campus UAB, 08193 Bellaterra
(Barcelona), Spain\
            \and Port d’Informació Científica (PIC), Campus UAB, C. Albareda s/n, 08193 Bellaterra (Barcelona), Spain\\ }

   \date{Received XX, 20XX}

 
  \abstract
  {Wide-field imaging surveys now provide photometry for billions of sources, while spectroscopic observations remain limited, motivating methods that can extract spectroscopic information from photometric data.}  
   {We present a generative framework for the joint probabilistic inference of galaxy redshifts and rest-frame spectra from broadband photometric fluxes. The model provides a sampling-based estimate of the photometric-redshift probability density function (PDF) for each galaxy, from which accurate point estimates are derived, and reconstructs rest-frame spectra that preserve key spectral properties.} 
   {We pre-train a spectral autoencoder, \textsc{spender}, on 5 million DESI DR1 spectra to learn a low-dimensional latent space that represents rest-frame spectra. Conditioned on galaxy broadband photometric fluxes, a diffusion model jointly infers the corresponding spectral latent representation and photometric redshift. The inferred latent representation is decoded into a high-resolution rest-frame spectrum, which can be transformed to the observed frame by redshifting and resampling.
}
   {Sampling from the conditional diffusion model yields a full photometric-redshift PDF for each galaxy, with the resulting point estimates showing a precision comparable to that of a gradient-boosted decision tree model. In most cases, the reconstructed rest-frame spectra reproduce the overall continuum shape and capture the presence of prominent spectral features. For galaxies with sufficiently high signal-to-noise ratios in their observed spectra, the Dn4000 index shows good agreement between the reconstructed spectra and the observed spectra. On average, the spectral reconstruction residuals are close to the noise level of the observed spectra.
}
   {Latent-diffusion generative modeling enables joint inference of galaxy photometric-redshift PDFs and rest-frame spectra from photometric fluxes.}

   \keywords{galaxies: distances and redshifts -- galaxies: photometry -- methods: data analysis}
\titlerunning{Joint probabilistic inference of galaxy redshifts and rest-frame
spectra with latent diffusion}
   \maketitle
\nolinenumbers

\section{Introduction}

Over the past decade, wide-field imaging surveys have grown dramatically in depth and area, producing catalogs that contain billions of astronomical sources and, consequently, vast samples of galaxies (e.g., see \citealt{2016MNRAS.460.1270D} for the Dark Energy Survey, \citealt{2016arXiv161205560C} for Pan-STARRS1, and \citealt{2019AJ....157..168D} for the DESI Legacy Imaging Surveys). Looking ahead, the Vera C.~Rubin Observatory Legacy Survey of Space and Time (LSST) is expected to observe $\sim$20 billion galaxies over its 10-year survey \citep{2019ApJ...873..111I}. At the same time, spectroscopic surveys have also expanded rapidly. For example, the Dark Energy Spectroscopic Instrument (DESI; \citealt{2016arXiv161100036D}) DR1 main survey \citep{Abdul_Karim_2026} includes spectra for more than 18 million unique targets. However, the total number of galaxy spectra remains orders of magnitude smaller than the number of photometric detections due to the substantially higher observational cost per object.

Spectroscopy provides an informative probe of galaxy properties. By analyzing specific spectral features, such as emission and absorption lines, researchers can accurately measure redshifts and derive physical properties such as star formation rates (SFRs) and metallicities \citep{1998ARA&A..36..189K,2004MNRAS.351.1151B,2002ApJS..142...35K}. To extend such analyses to large photometric samples, the standard approach relies on fitting multi-band photometry to theoretical spectral energy distributions (SEDs) \citep{2011Ap&SS.331....1W}. While enabling the analysis of large samples, these methods are limited by the adopted SED model space and priors, which can restrict the flexibility of the inferred galaxy properties and introduce degeneracies and systematic uncertainties \citep{2013ARA&A..51..393C}.

Data-driven and machine-learning (ML) approaches have emerged as a powerful alternative to bridge the gap between photometric and spectroscopic surveys. Unlike traditional template-fitting methods, these approaches exploit correlations in large datasets to infer galaxy information directly from photometric observations. They have been widely used to predict galaxy redshifts and, more recently, to reconstruct galaxy spectra from photometry.
For example, \citet{2020arXiv200912318W} used the autoencoder framework of \citet{2020AJ....160...45P} to reconstruct galaxy spectra from broadband images. This framework was applied by \citet{2021ApJ...914..142H} to the giant spiral galaxy UGC 2885, demonstrating the capacity of ML models to infer spectral features. Later, \citet{2022arXiv221105556D} adapted the Denoising Diffusion Probabilistic Model (DDPM) framework of \citet{ho2020denoising} to reconstruct observed-frame spectra from broadband galaxy images using a conditional diffusion model. Subsequently, \citet{2024ApJ...977..131D} extended this line of work by carrying out spectroscopic analyses of the spectra generated by this approach. Their analysis showed that the reconstructed spectra are not only visually consistent with observations, but also retain sufficient physical fidelity to allow for the derivation of key galaxy properties. Collectively, these works show that generative models can recover realistic spectral information from photometric data.

However, machine-learning approaches that directly infer galaxy properties from photometric data usually treat redshift estimation and spectral reconstruction as separate tasks. This contrasts with template-fitting methods, which typically explore redshift and SED space jointly by forward-modeling photometric fluxes from redshifted templates. For broadband photometric observations, redshift inference is closely linked to the underlying galaxy spectrum, since degeneracies in color–redshift space reflect ambiguities in the corresponding spectral energy distributions. Modeling this joint structure can therefore help provide more consistent and interpretable inference.

In this work, we present a generative framework for joint probabilistic inference of galaxy redshifts and rest-frame spectra from photometric fluxes. Our approach models the joint conditional distribution \(p(z,\mathbf{s}\mid\mathbf{X})\), where \(z\) is the redshift, \(\mathbf{s}\) is a latent representation of the rest-frame spectrum, and \(\mathbf{X}\) denotes the input photometric fluxes. This formulation enables redshift and spectral information to be inferred jointly, while allowing joint and potentially non-Gaussian uncertainties to be represented through sampling. The sampled spectral latent representations are then decoded into rest-frame spectra, which can subsequently be transformed to the observed frame through redshifting and resampling.

Our method follows the latent-diffusion paradigm \citep{rombach2022high} by combining a pre-trained spectral autoencoder, \textsc{spender} \citep{2023AJ....166...74M}, with a conditional diffusion model. The diffusion model performs generative modeling in the compact spectral latent space defined by the pre-trained \textsc{spender} autoencoder, rather than directly in the high-dimensional spectral-pixel space, thereby reducing computational cost while maintaining generative fidelity \citep{rombach2022high}. In \textsc{spender}, the encoder maps an observed-frame spectrum to a compact latent representation, while the decoder outputs a rest-frame spectrum. The decoded rest-frame spectrum is mapped to the observed frame through an explicit analytical redshift transformation followed by resampling, rather than being learned implicitly by the decoder. This design encourages the latent representation to encode intrinsic rest-frame spectral information rather than redshift effects.

We train and evaluate our method on the DESI DR1 dataset \citep{Abdul_Karim_2026}, which
provides a very large, uniformly processed spectroscopic sample with reliable redshifts and wide coverage in galaxy type and redshift.
Compared with earlier image-based reconstructions \citep{2020arXiv200912318W, 2022arXiv221105556D}, our approach leverages a larger spectroscopic training set, spans a broader redshift range, and operates directly on photometric fluxes rather than image data.
Although images can provide additional spatial or morphological information, using photometric fluxes simplifies the pipeline and facilitates its application to other surveys based on standard multi-band measurements, while still exploiting the color information that encodes spectral variation.

The remainder of this paper is organized as follows. 
Sect.~\ref{sec:data} describes the DESI DR1 spectroscopy and the matched photometry used in this work. 
Sect.~\ref{sec:method} introduces our reconstruction pipeline, including the model architecture and the training procedure.
Sect.~\ref{sec:results} presents qualitative and population-level results for spectral reconstruction and photometric redshifts, together with diagnostics based on continuum indices. 
Sect.~\ref{sec:summary} summarizes our findings and discusses limitations and future directions.

\section{Data}\label{sec:data}

\subsection{DESI DR1 spectra}\label{data:spectra}
The spectra used in this work are from the DESI DR1 \citep{Abdul_Karim_2026}. 
The DR1 data release comprises all data from the first 13 months of the main survey, together with a uniform reprocessing of the Survey Validation data previously included in the DESI Early Data Release. The DESI DR1 main survey provides high-confidence redshifts for 18.7 million objects, of which 13.1 million are spectroscopically classified as galaxies, making DESI DR1 the largest extragalactic redshift sample to date. 
DESI spectra cover 3600–9824\,\AA{} with a sampling of 0.8\,\AA{} per pixel. In this work we use the DESI DR1 main survey data, which comprise more than 90\% of the useful spectra in DR1 and are divided into three observing programs: \texttt{dark}, \texttt{bright}, and \texttt{backup}.

We adopt the main-survey target-coadded spectra (HEALPixel-based) together with the corresponding redshift catalogs.
To construct a reliable galaxy sample, we select objects by applying the same set of quality cuts to these redshift catalogs of each observing program. Guided by the data selection strategy of \citet{2024A&A...691A.308S}, we require that the catalog entry provides the recommended redshift in each observing program (\texttt{ZCAT\_PRIMARY}=\texttt{TRUE}), ensure the coadded spectra have no fiber issues (\texttt{COADD\_FIBERSTATUS}=0), and demand a reliable redshift (\texttt{ZWARN} = 0 or 4). Our sample is restricted to sources classified as galaxies (\texttt{SPECTYPE}=\texttt{'GALAXY'}). In addition, we impose a redshift-uncertainty cut of $\texttt{ZERR} < 5\times10^{-4}$ and require redshifts greater than 0.01 to reduce possible stellar contamination.
We then retrieve the spectra using the unique \texttt{TARGETID} identifiers, yielding a total sample of approximately 14.6 million spectra.

We randomly select 5 million spectra from this sample for training \textsc{spender}. Following the procedure described by \citet{2023AJ....166...74M}, we apply a mask to atmospheric emission lines by setting the inverse-variance weights (\texttt{ivar}) to zero at the affected wavelengths. Consequently, all downstream calculations, including training loss computation, S/N estimation, and reconstruction metrics, are restricted to valid pixels where $\texttt{ivar} > 0$, ensuring that the results are not affected by masked regions.

The spectra used in this work extend to redshifts of about 1.7, substantially beyond the redshift range considered by \citet{2023AJ....166...74M}, whose sample selection is limited to about 0.5. We therefore adopt an amplitude normalization scheme that remains valid over this broader redshift range. In \citet{2023AJ....166...74M}, each spectrum is normalized by the median flux measured within a fixed rest-frame wavelength window. This requires that the chosen window remain within the observed-frame wavelength coverage. Once it shifts outside that range, the normalization can no longer be applied. In practice, this limits the usable sample to lower redshift.
To avoid this limitation, we normalize each spectrum by the median flux over its full available wavelength coverage, ensuring that the normalization is applicable across the full redshift range. During \textsc{spender} pre-training, we apply random multiplicative rescalings to encourage the model to focus on spectral shape rather than global amplitude (see Sect.~\ref{sec:Spender}).

\begin{figure}[!t] 
  \centering
  \includegraphics[width=\columnwidth]{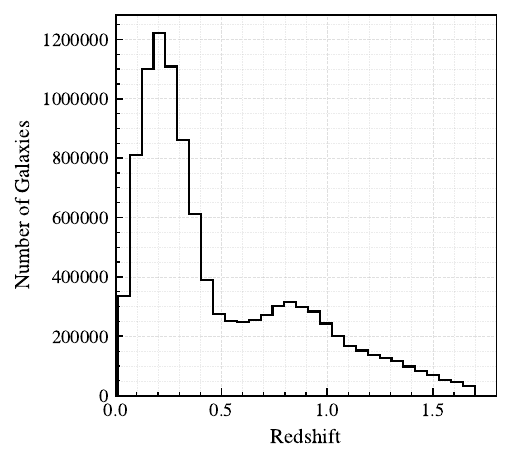}
  \caption{Redshift distribution of approximately 10.5 million galaxies in the final sample used in this work. The sample is constructed after removing galaxies with multiple coadded spectra and applying the photometric magnitude cuts.}
  \label{fig:z_dist}
\end{figure}

\subsection{Photometry}\label{sec:Photometry}

The photometric fluxes used in this work are from Data Release 9 of the DESI Legacy Imaging Surveys \citep{2019AJ....157..168D}, which provides optical $g$, $r$, and $z$ bands alongside mid-infrared $W1$ and $W2$ photometry derived from the WISE mission \citep{2010AJ....140.1868W}. The $g,r,z,W1$ and $W2$ bands provide color information that constrains galaxy SED shapes and has been used for photometric-redshift prediction based on DESI Legacy Surveys imaging \citep{2021MNRAS.501.3309Z}. These photometric measurements are available for DESI spectroscopic targets and are already matched to the spectra in the DESI DR1 data products.

Our goal is to reconstruct spectra from photometric fluxes and benchmark the reconstructions against observed spectra, which requires a one-to-one association between each photometric input and a corresponding observed spectrum. To achieve this, after training \textsc{spender}, we removed the small fraction of galaxies (approximately 1\%) containing multiple coadded spectra, ensuring each galaxy is associated with a unique spectrum.
To ensure high photometric quality, we converted fluxes to AB magnitudes and applied cuts of $g \le 24.0$, $r \le 23.4$, and $z \le 22.5$. These thresholds correspond to the approximate $5\sigma$ depth of the DESI Legacy Imaging Surveys \citep{2019AJ....157..168D}, effectively excluding measurements with low signal-to-noise ratios (S/N).
The final sample comprises approximately 10.5 million galaxies. As shown in Fig.~\ref{fig:z_dist}, the selected galaxies span a broad redshift range, extending up to $z \approx 1.7$.

For the data split, all galaxies used for \textsc{spender} pre-training and retained in the final sample were assigned to the training partition. The remaining galaxies were then randomly split so that the final training, validation, and test sets followed a $70\%{:}15\%{:}15\%$ split. All reported metrics are computed on held-out test galaxies that were used neither for \textsc{spender} pre-training nor for training the conditional diffusion model.

\section{Method}\label{sec:method}

In this section, we first present the overall pipeline for joint redshift and spectrum inference in Sect.~\ref{sec:pipeline}. We then describe its main technical components: Sect.~\ref{sec:Spender} summarizes the \textsc{spender} autoencoder, and Sect.~\ref{sec:diffu} details the conditional diffusion model.

\begin{figure*}[!t] 
  \centering
  \includegraphics[width=\textwidth]{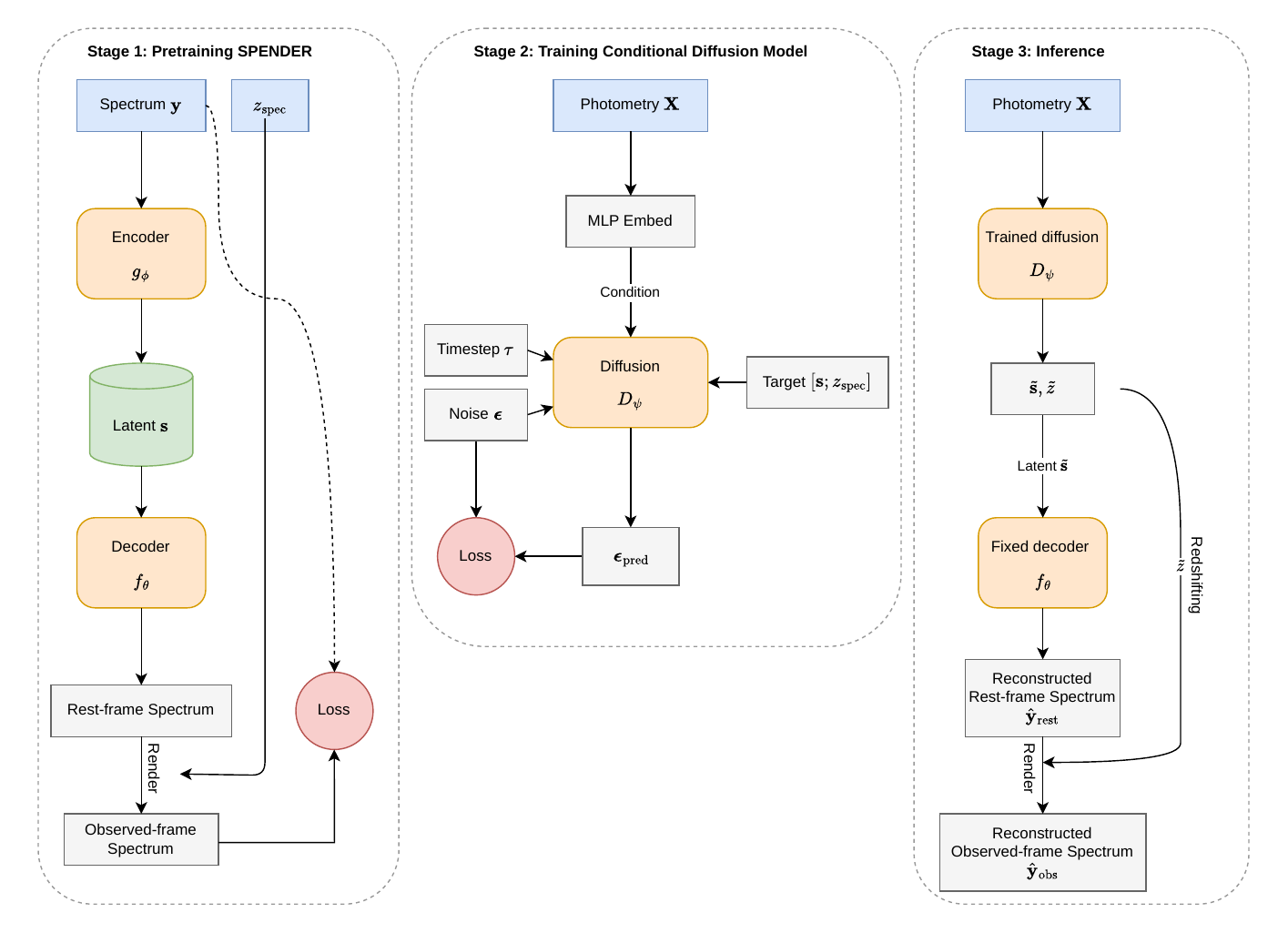}
  \caption{Schematic overview of the method. In Stage 1, the \textsc{spender} autoencoder learns a compact latent representation $\mathbf{s}$ from spectra. In Stage 2, a conditional diffusion model $D_\psi$ is trained to predict $\mathbf{s}$ and redshift from photometric fluxes $\mathbf{X}$. At inference time, the pipeline jointly infers redshifts and spectra from photometric fluxes alone.}
  \label{fig:method_pipeline}
\end{figure*}

\subsection{Overall pipeline for joint redshift and spectrum inference}\label{sec:pipeline}

Our method infers galaxy redshifts and spectra from photometry through a learned latent representation. The pipeline consists of two training stages followed by an inference phase, as illustrated in Fig.~\ref{fig:method_pipeline}.

Let $\mathbf{X} \in \mathbb{R}^5$ denote the input flux measurements (in the $g,r,z,W1,W2$ bands), and let $\mathbf{y}\in\mathbb{R}^{M}$ denote the normalized observed-frame spectrum sampled on an $M$-pixel wavelength grid, with corresponding inverse-variance weights $\mathbf{w}\in\mathbb{R}^{M}$. 
We denote $z_{\rm spec}$ and $z_{\rm phot}$ as the spectroscopic and photometric redshifts, respectively. The core of our method relies on a latent vector $\mathbf{s}\in\mathbb{R}^{S}$, which serves as a compact representation of the galaxy rest-frame spectrum.

In the first stage, we pre-train the spectral autoencoder \textsc{spender}, which consists of an encoder $g_\phi$ and a decoder $f_\theta$. The encoder $g_\phi$ takes the observed-frame spectrum $\mathbf{y}$ as input and produces a compact, low-dimensional representation $\mathbf{s}$ of the galaxy's rest-frame spectrum.
The decoder $f_\theta$ reconstructs the rest-frame spectrum from $\mathbf{s}$. This spectrum is then shifted back to the observed frame via an analytical redshift transformation  (i.e. not a learned network operation) based on the spectroscopic redshift
and resampled onto the DESI wavelength grid before computing the loss. 
Further details of the \textsc{spender} pre-training are provided in Sect.~\ref{sec:Spender}. 

In the second stage, we train a conditional diffusion model, $D_\psi$, to predict the spectral latent vector and redshift from the photometric fluxes. 
We treat the target as the concatenation $\mathbf{t}=[\mathbf{s}; z_{\rm spec}]$ and condition the model on the photometric fluxes $\mathbf{X}$ to learn the conditional distribution $p(\mathbf{t}\,|\,\mathbf{X})$. 
Further details of the diffusion model are given in Sect.~\ref{sec:diffu}.

At inference time, the trained diffusion model generates samples of the spectral latent vector and redshift, $(\tilde{\mathbf{s}}, \tilde{z})$, given only the photometric input $\mathbf{X}$. Each sampled latent vector is decoded into a rest-frame spectrum using the trained \textsc{spender} decoder, and the resulting spectrum can be mapped to the observed frame via a redshift transformation and resampling onto the desired wavelength grid.

\subsection{Spectral autoencoder (\textsc{spender})}\label{sec:Spender}

We utilize \textsc{spender} \citep{2023AJ....166...74M}, a deep-learning framework designed to encode galaxy spectra into a low-dimensional latent space that captures their intrinsic rest-frame spectral characteristics. The encoder of \textsc{spender} is based on a convolutional neural network (CNN) with an attention mechanism that helps the model identify informative spectral features in the observed-frame spectrum, where their locations vary with redshift. The encoder compresses the input spectrum $\mathbf{y}$ into a spectral latent vector $\mathbf{s}$, while the decoder maps $\mathbf{s}$ to a rest-frame spectrum. The transformation from the rest frame to the observed frame is not learned implicitly by the network. 
Instead, it is implemented explicitly through a deterministic redshift-and-resampling operation. During training, the decoded rest-frame spectrum is redshifted using the known spectroscopic redshift $z_{\rm spec}$ and resampled onto the observed-frame wavelength grid for direct comparison with the input spectrum $\mathbf{y}$. This design is intended to encourage the latent representation to focus on rest-frame spectral information rather than on the effects of redshift. This framework has also been applied to DESI Bright Galaxy Survey spectra for unsupervised outlier detection, demonstrating that it can learn useful latent representations from DESI data as well \citep{2023ApJ...956L...6L}. \citet{2023AJ....166...74M} showed that high-fidelity spectral reconstruction can be achieved with a latent dimensionality of $S=6$--$10$. We therefore adopt a latent dimensionality of \(S=10\).

We train \textsc{spender} on DESI DR1 spectra sampled on the observed-frame wavelength grid ($3600$--$9824$~\AA{}, with 7,781 pixels). The maximum redshift in our DESI DR1 spectral sample is close to $1.7$, and we therefore adopt $z_{\max}=1.7$. We construct a linear rest-frame wavelength grid spanning from the de-redshifted lower limit $\left(3600/(1+z_{\max})\right)\,\text{\AA}$ to $9824\,\text{\AA}$. The number of rest-frame grid points is set to approximately $(1+z_{\max})$ times that of the observed-frame grid, following the design principle of \textsc{spender}. This choice avoids undersampling in the subsequent resampling step.

To obtain latent representations that are approximately invariant to redshift variation, we adopt the training strategy proposed by \citet{2023AJ....166...75L}. The training objective $\mathcal{L}_{\rm total}$ combines a fidelity term $\mathcal{L}_{\rm fid}$ with two regularization terms:
\begin{equation}
\mathcal{L}_{\rm total} = \mathcal{L}_{\rm fid} + \mathcal{L}_{\rm sim} + \mathcal{L}_{\rm c}.
\end{equation}
The fidelity loss $\mathcal{L}_{\rm fid}$ is the weighted mean squared error between the input spectrum and the reconstructed observed-frame spectrum. The similarity loss $\mathcal{L}_{\rm sim}$ regularizes the latent space by encouraging distances in latent space to be consistent with distances between the corresponding rest-frame spectra. The consistency loss $\mathcal{L}_{\rm c}$ reduces the distance between the latents of an original spectrum and its augmented counterpart.
We adopt the original augmentation scheme with two modifications. First, instead of drawing a new redshift uniformly over the full training range, we perturb the original redshift with a Gaussian scatter of width $\sigma_z=0.1$. Second, we apply a random multiplicative amplitude jitter of $10\%$ to the augmented spectra and enforce latent consistency under this perturbation. This encourages the encoder to reduce its sensitivity to global amplitude variations and to focus more strongly on spectral shape.
Minimizing the combined objective yields a latent space that is approximately invariant to redshift and global amplitude scaling, providing stable latent targets for the subsequent diffusion model training.

We pre-trained \textsc{spender} in a distributed configuration using six NVIDIA L40S GPUs. Computations were carried out in bfloat16 precision, and parallelization across devices was handled using the Hugging Face Accelerate framework. To ensure efficient data throughput, spectra were loaded directly to GPU memory through a custom data-loading pipeline. The loader reads contiguous blocks of 10,000 spectra from disk and iterates over these blocks in mini-batches during training, providing sufficient stochasticity while minimizing I/O overhead. We further enforced an identical number of batches on each GPU to avoid synchronization issues during distributed optimization. The model was trained for ten epochs using a one-cycle learning-rate schedule with a maximum learning rate of $10^{-3}$ \citep{smith2019super}, taking approximately 8--9 hours on the six-GPU setup. The trained \textsc{spender} encoder is subsequently used to generate ground-truth latent vectors as a preprocessing step for downstream analyses, while the decoder is kept fixed and used to map latent vectors to rest-frame spectra.

\subsection{Conditional diffusion model}\label{sec:diffu}

After the pre-training of \textsc{spender}, we employ a conditional diffusion model, following the DDPM formulation of \citet{ho2020denoising}, to learn the probabilistic mapping from photometry to the spectral latent vector and redshift. For the training sample, we use the fixed \textsc{spender} encoder $g_{\phi}$ to deterministically encode each observed spectrum into a spectral latent vector $\mathbf{s}$, and concatenate it with the corresponding spectroscopic redshift to define the supervision vector
$\mathbf{t}_0 \in \mathbb{R}^{S+1}$:
$\mathbf{t}_0=[\mathbf{s};z_{\rm spec}]$.
The diffusion model is trained to learn $p_{\theta}(\mathbf{t}_0 \mid \mathbf{X})$, where $\mathbf{X}=(g,r,z,W1,W2)$ denotes the input broadband fluxes. Compared with direct point estimation, this framework provides a flexible way to model the full conditional distribution $p_{\theta}(\mathbf{t}_0 \mid \mathbf{X})$, and thus enables uncertainty to be quantified via samples from the conditional distribution.

The diffusion model defines a generative process through two stages: a forward process that progressively adds Gaussian noise to the target vector, and a learned reverse process that iteratively removes noise conditioned on the input photometry $\mathbf{X}$. In the forward process, Gaussian noise is added to $\mathbf{t}_0$ over $T$ steps according to a variance schedule $\beta_1,\dots,\beta_T$, giving
\begin{equation}
\mathbf{t}_\tau = \sqrt{\bar{\alpha}_\tau}\,\mathbf{t}_0 + \sqrt{1-\bar{\alpha}_\tau}\,\boldsymbol{\epsilon},
\qquad
\boldsymbol{\epsilon}\sim\mathcal{N}(\mathbf{0},\mathbf{I}),
\end{equation}
where $\alpha_\tau = 1-\beta_\tau$ and $\bar{\alpha}_\tau=\prod_{s=1}^{\tau}\alpha_s$. We adopt a linear noise schedule with $\beta_\tau \in [10^{-4},\,2\times10^{-2}]$ and $T=200$.
In the reverse process, a neural network $\boldsymbol{\epsilon}_\theta$ is trained to predict the injected noise from the noisy target state $\mathbf{t}_\tau$, the diffusion timestep $\tau$, and the conditioning photometric fluxes $\mathbf{X}$. The conditioning vector $\mathbf{X}$ is encoded by a dedicated multilayer perceptron (MLP), while the timestep is represented using a sinusoidal embedding \citep{vaswani2017attention}. 

The model is optimized with the standard DDPM noise-prediction objective, minimizing the mean-squared error between the true Gaussian noise injected in the forward process and the noise predicted by the network, with diffusion timesteps sampled uniformly during training. Projecting spectra into the low-dimensional \textsc{spender} latent space reduces the dimensionality of the target representation while retaining the dominant spectral information. In this compact space, the conditional diffusion model can flexibly represent the joint conditional distribution of the spectral latent vector and redshift, including non-Gaussian or potentially multimodal prediction uncertainties arising from photometric degeneracies.

We train the model using the AdamW optimizer \citep{loshchilov2018decoupled} with a learning rate of $10^{-4}$, a weight decay of $10^{-4}$, and a batch size of 1024. The model is trained for up to 50 epochs, with early stopping based on the root mean squared error on the validation set and a patience of ten epochs. We maintained an exponential moving average (EMA) of the model parameters and used the EMA model for sampling at inference time.

At inference time, only the photometric input $\mathbf{X}$ is required. We generate samples from the conditional distribution $p_\theta(\mathbf{t}_0\mid\mathbf{X})$ using the Denoising Diffusion Implicit Models (DDIM) sampler \citep{song2020denoising}, reducing the reverse process from 200 to 50 denoising steps. The training of the diffusion model was completed in approximately 30 minutes on a single NVIDIA L40S GPU. For each galaxy, we draw $N=200$ samples from $p_\theta(\mathbf{t}_0 \mid \mathbf{X})$, which are used for downstream analyses. For the approximately 1.5 million galaxies in the test set, generating 200 samples per galaxy takes about 3 hours on our computing setup, corresponding to a throughput of approximately 140 galaxies per second. The implementation of the conditional diffusion model, along with additional architectural details, is publicly available.\footnote{\url{https://github.com/hanyguo/conditional-diffusion}}

\begin{figure*}[!t]\label{sec:corner}
\centering
\includegraphics[width=\textwidth]{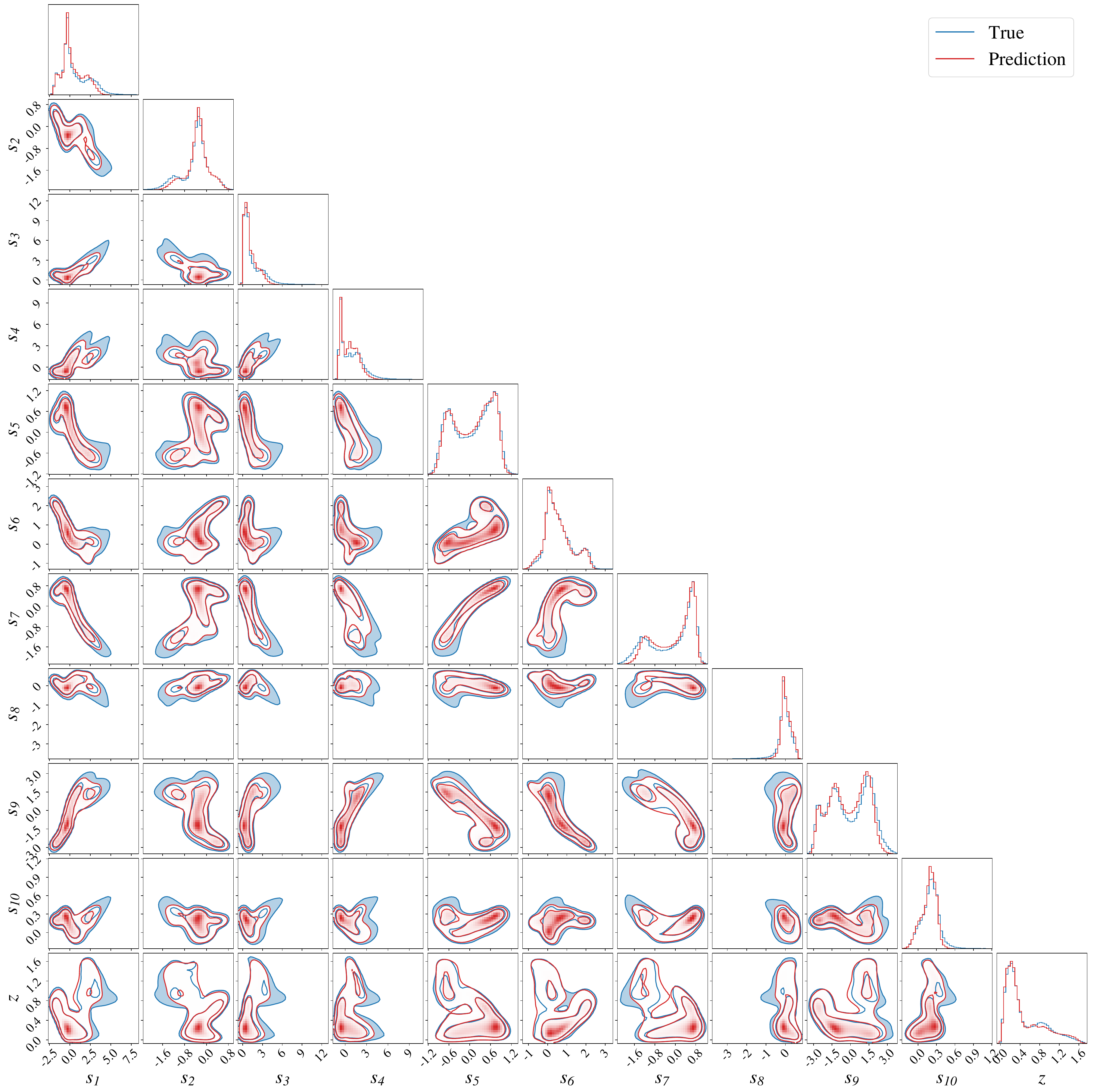}  
\caption{Comparison between the ground-truth distribution (blue) and the diffusion model predictions (red) over the full test set. The distributions cover the \textsc{spender} latent vectors and redshifts. For each galaxy, the predicted distribution (red) is formed by drawing one sample from the conditional diffusion model. The diagonal panels present the 1D marginalized distributions of each variable. Meanwhile, the off-diagonal panels show the corresponding 2D projections with 68\% and 95\% density contours. Axis limits in each panel are set by the 0.1\% and 99.9\% quantiles of the combined distributions.}
\label{fig:corner}
\end{figure*}

\begin{figure*}[!t] 
\centering
\includegraphics[width=\textwidth]{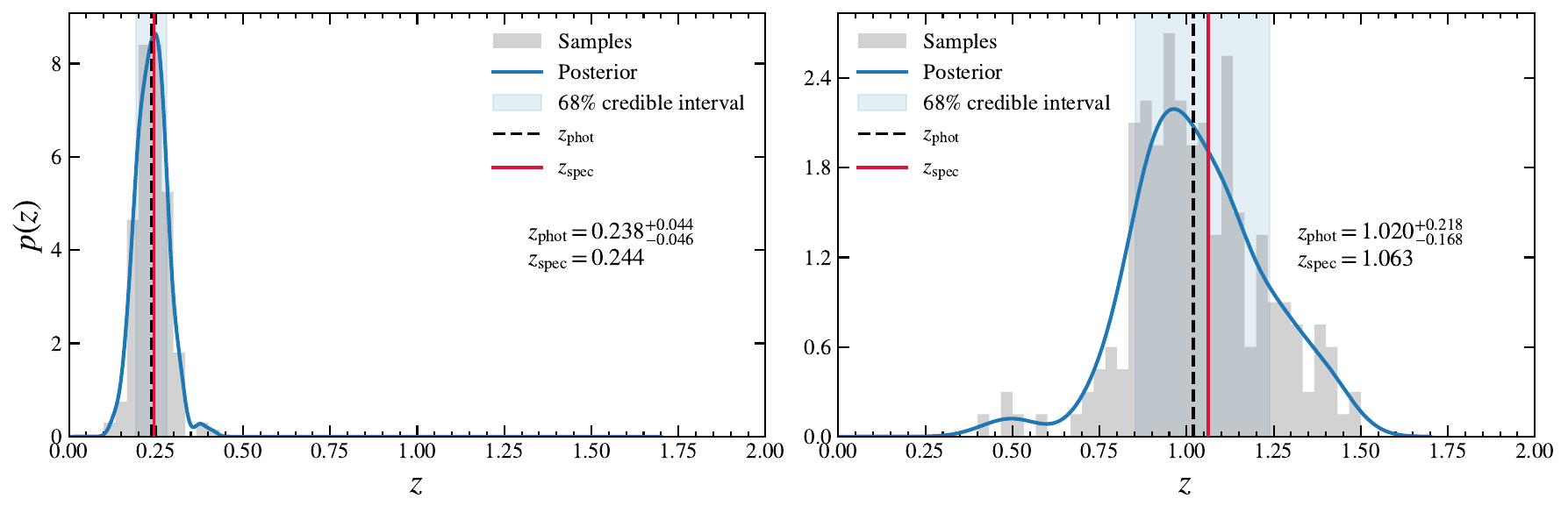}
\caption{Posterior redshift distributions predicted by the conditional diffusion model for a narrow case (left; TARGETID: 39633467100627984) and a broad case (right; TARGETID: 39627702818312612). The inferred redshift PDF is shown as a blue curve, estimated from the ensemble of sampled predictions using kernel density estimation (KDE), while the gray histogram indicates the distribution of the prediction samples.
The black dashed vertical line marks the point estimate $z_{\rm phot}$ (the posterior median), and the red solid vertical line indicates the spectroscopic redshift $z_{\rm spec}$.
The light--blue shaded region denotes the 68\% credible interval (16th--84th percentiles). }
\label{fig:pdf}
\end{figure*}

\begin{figure*}[!t] 
 \centering
  \includegraphics[width=\textwidth]{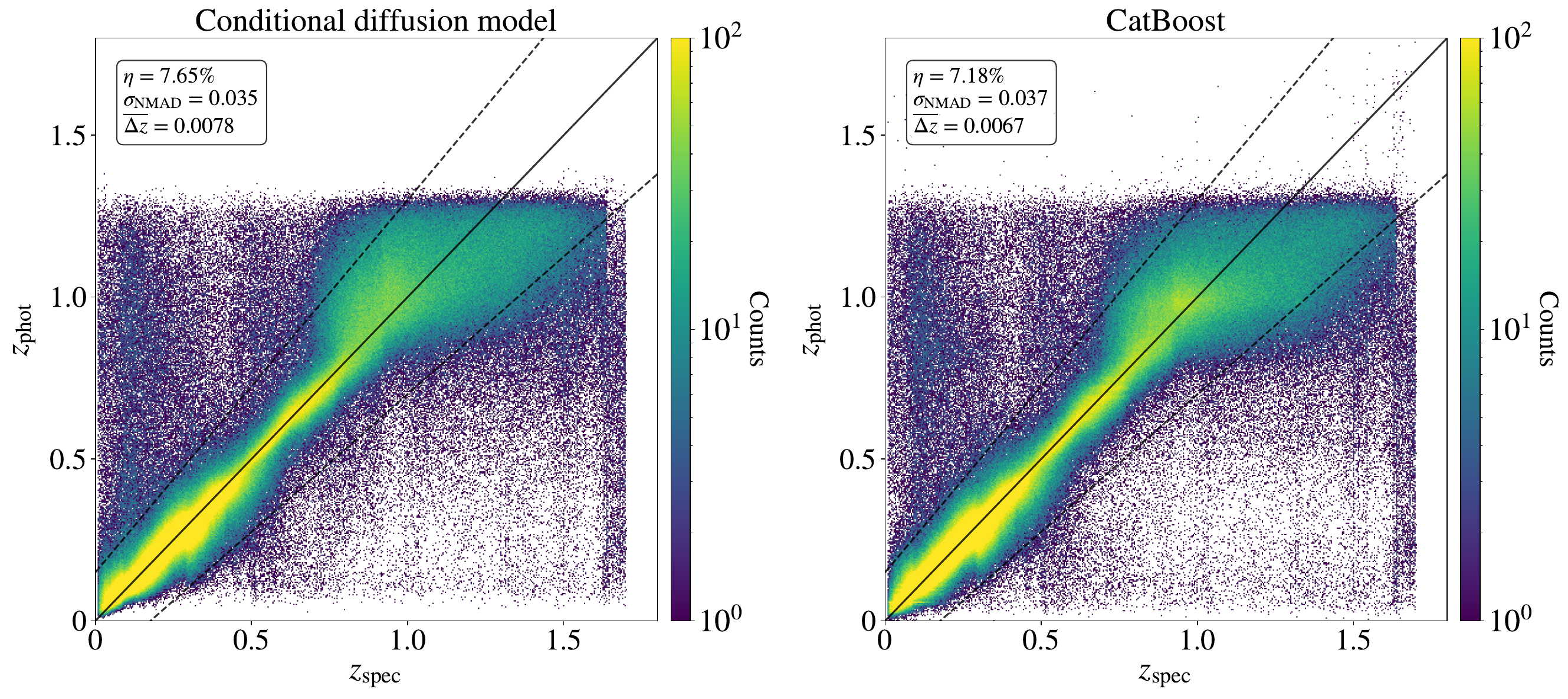}
  \caption{Scatter plot of spectroscopic redshift ($z_{\rm spec}$) versus photometric redshift ($z_{\rm phot}$) for the test set. The left panel shows the conditional diffusion model predictions, while the right panel shows results from CatBoost. The quantities $\eta$, $\sigma_{\rm NMAD}$, and $\langle \Delta z \rangle$ denote the outlier fraction, the normalized median absolute deviation ($\sigma_{\mathrm{NMAD}}$), and the mean bias, respectively, where $\Delta z \equiv (z_{\rm phot}-z_{\rm spec})/(1+z_{\rm spec})$. The black solid line indicates the one-to-one relation, while the black dashed lines mark the outlier threshold at $|\Delta z| = 0.15$. The color bar shows the source density per pixel.}
  \label{fig:photoz}
\end{figure*}

\begin{figure}[!t] 
  \centering
  \includegraphics[width=\columnwidth]{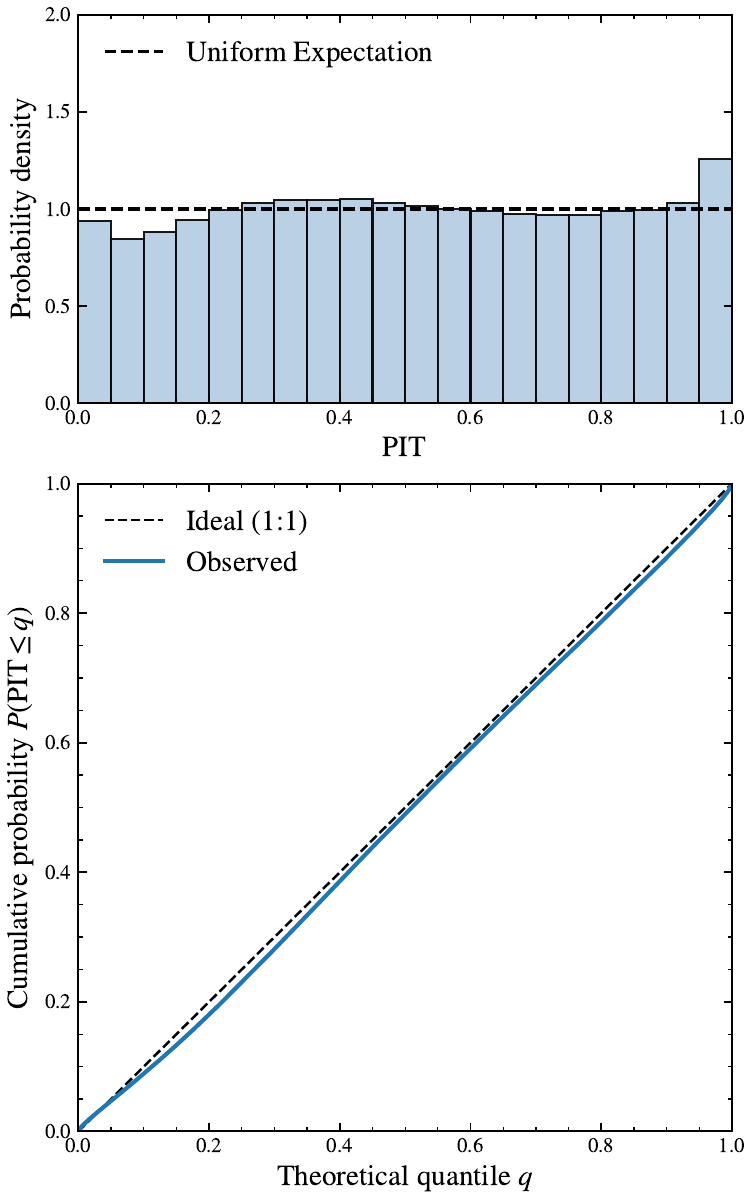}
  \caption{Probabilistic calibration diagnostics for the photometric-redshift posteriors from the conditional diffusion model on the test set. 
\emph{Top:} The probability integral transform (PIT) histogram. The dashed line indicates the uniform distribution expected for perfectly calibrated posteriors. 
\emph{Bottom:} Probability--probability (P--P) plot comparing the empirical cumulative distribution function (CDF) of the PIT values (solid line) with the ideal one-to-one relation (dashed line).}
  \label{fig:pit}
\end{figure}

\section{Results}\label{sec:results}

\subsection{Validating the spectral autoencoder}

We pre-trained \textsc{spender} on 5 million spectra selected from DESI DR1 (Sect.~\ref{data:spectra}) and evaluate its reconstruction performance on the test set. For each galaxy, the encoder maps the observed spectrum to a latent vector, from which the decoder reconstructs the corresponding rest-frame spectrum. This reconstructed rest-frame spectrum is then redshifted and resampled onto the observed-frame wavelength grid for direct comparison with the input observed spectrum.

For each spectrum, we quantify the reconstruction residual using the mean chi-square per pixel,
\begin{equation}
    \chi^2/N \;=\; \frac{1}{N}\sum_{i=1}^{N} w_i\left(y_{i}^{\rm obs} - y_{i}^{\rm recon}\right)^2,
\label{eq:chi2_per_pixel}
\end{equation}
where $y_{i}^{\rm obs}$ and $y_{i}^{\rm recon}$ are the observed and reconstructed fluxes at spectral pixel $i$, $w_i = 1/\sigma_i^2$ is the inverse-variance weight at the $i$th pixel, and the summation runs over the $N$ valid spectral pixels with $w_i>0$.
A distribution of $\chi^2/N$ centered near unity indicates that the reconstruction residuals are broadly consistent with the observational uncertainties.
We evaluate this metric on a subset of 10{,}000 galaxies uniformly sampled from the test set (hereafter the ``10k reconstruction sample'') to assess reconstruction fidelity at the population level. For each galaxy in this sample, we first compute $\chi^2/N$ from Eq.~(\ref{eq:chi2_per_pixel}), and then average the resulting values over the full sample. We obtain $\langle \chi^2/N \rangle = 1.038$, indicating that the residuals are, on average, consistent with the observational uncertainties.
This suggests that \textsc{spender} compresses the spectra into a latent representation that captures the dominant spectral information while maintaining accurate reconstructions.
This section evaluates only the autoencoder reconstruction from spectroscopic inputs, whereas Sects.~\ref{sec:rest-frame spectra} and~\ref{sec:obs-frame spectra} assess the full photometry-to-spectrum pipeline, in which the latent vector is inferred from photometry rather than encoded from the observed spectrum. These sections also present representative reconstructed spectra.

\subsection{Latent-space inference from photometry}
\label{sec:results_prediction}

As described in Sect.~\ref{sec:pipeline}, our framework jointly predicts the spectral latent vector and the redshift from the photometric fluxes $\mathbf{X}$. We perform this using the conditional diffusion model described in Sect.~\ref{sec:diffu}. For a given galaxy with photometry $\mathbf{X}$, the model defines the conditional joint distribution $p(\mathbf{s}, z \mid \mathbf{X})$. A single draw from this distribution yields one predicted pair $(\tilde{\mathbf{s}}, \tilde{z})$, while repeated draws produce a set of samples $\{(\tilde{\mathbf{s}}^{(k)}, \tilde{z}^{(k)})\}_{k=1}^{N_{\rm samp}}$ that approximates the joint conditional distribution.

Fig.~\ref{fig:corner} compares the joint distribution of the latent variables and redshift predicted by the conditional diffusion model with the corresponding target distribution on the test set. For each test-set galaxy, we draw one sample from the conditional diffusion model. Collecting these samples across the test set yields a predicted distribution (red), which we compare to the corresponding (ground-truth) distribution of latent vectors encoded from the spectra and spectroscopic redshifts (blue).
The diagonal panels show the one-dimensional marginalized distributions of each latent variable and the redshift, while the off-diagonal panels display the corresponding two-dimensional projections with 68\% and 95\% density contours. The predicted samples show qualitative agreement with the ground truth in both their marginal distributions and their main pairwise structure.
This comparison provides a population-level check that photometry-conditioned sampling yields latent representations occupying similar regions of the \textsc{spender} latent space as those derived from spectra. The quality of the spectra reconstructed from these latent representations is evaluated in Sect.~\ref{sec:rest-frame spectra}.

\begin{figure*}[!t]
\centering
\includegraphics[width=\textwidth]{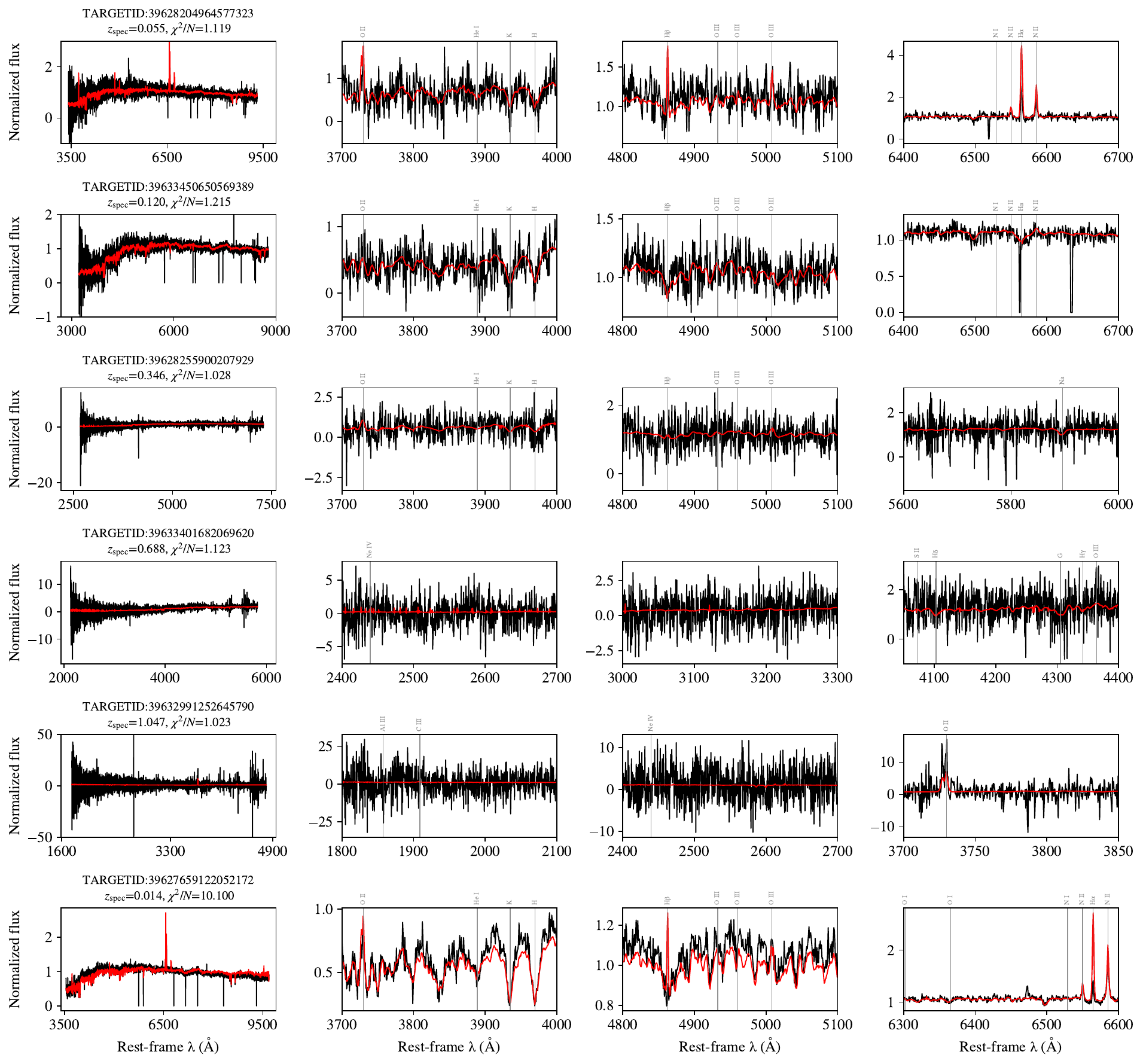}  
\caption{Observed (black) and reconstructed (red) spectra for six representative galaxies in the rest frame. The first five rows show examples randomly selected from five spectroscopic-redshift intervals and ordered by increasing spectroscopic redshift. 
The sixth row shows an example with a high reconstruction residual, randomly selected from the 1\% of objects with the largest \(\chi^2/N\) values. The left column displays the full rest-frame spectrum for each galaxy, with the panel title indicating the \texttt{TARGETID}, spectroscopic redshift $z_{\rm spec}$, and $\chi^2/N$. The three right columns show zoomed-in spectral regions with annotated emission and absorption features. }
\label{fig:rest-frame_spectra}
\end{figure*}

\begin{figure*}[!t]
\centering
\includegraphics[width=\textwidth]{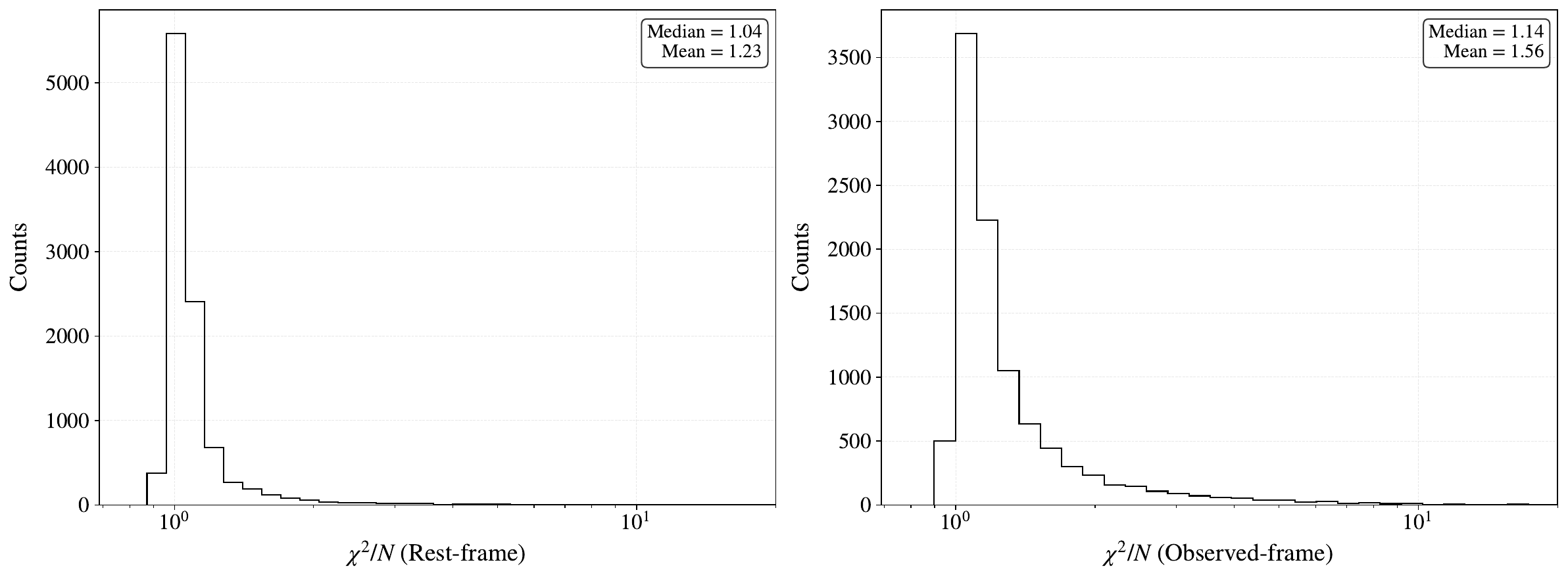}  
\caption{Distributions of spectral reconstruction residuals, quantified by the mean chi-square per pixel, \(\chi^{2}/N\), for the 10k reconstruction sample. Left: residuals measured in the rest frame. Right: residuals measured in the observed frame. Insets show the median and mean values in each panel.}
\label{fig:chiqdist}
\end{figure*}

\subsection{Photometric redshift estimation}
\label{sec:photoz_prediction}

We used the trained conditional diffusion model to predict the photometric redshift $z_{\rm phot}$ from photometric fluxes.
Instead of producing a single point estimate, diffusion sampling yields a redshift distribution for each galaxy, allowing non-Gaussian uncertainties to be represented.
The conditional diffusion model defines a joint distribution over redshift and spectral latent vector, conditioned on the photometric fluxes $\mathbf{X}$. Sampling this distribution yields paired draws $(z,\mathbf{s})$, where $\mathbf{s}$ denotes the latent representation of the rest-frame spectrum in the \textsc{spender} latent space. This joint inference is analogous to template-fitting photometric-redshift methods, which infer a joint distribution over redshift and spectral energy distributions (SEDs) by matching photometry to template SEDs. In contrast, many non-template approaches return only $p(z)$ without an explicit representation of the corresponding SEDs, which limits the interpretability of the inferred redshifts.
In our framework, the inferred SED is represented by the latent vector $\mathbf{s}$ in a low-dimensional latent space.

For each galaxy, we draw $N_{\rm samp}=200$ samples from the conditional diffusion model and estimate a smooth redshift PDF, $p(z)$, using Gaussian kernel density estimation (KDE) implemented in \texttt{scipy.stats.gaussian\_kde} with its default bandwidth selection \citep{virtanen2020scipy}.
We adopt the posterior median as the point estimate of the photometric redshift, denoted $z_{\rm phot}$, and report the 68\% credible interval defined by the 16th and 84th percentiles of $p(z)$. Fig.~\ref{fig:pdf} shows two representative posterior redshift distributions. The gray histogram represents the sampled redshift predictions, the blue curve shows the KDE-estimated PDF, the black dashed line marks $z_{\rm phot}$, the red line indicates the spectroscopic redshift $z_{\rm spec}$, and the shaded region denotes the 68\% credible interval. This distributional output is a key benefit of the conditional diffusion approach, as it provides both a point estimate and a full predictive distribution that can capture non-Gaussian or potentially multimodal redshift uncertainties.

We evaluate photometric redshift performance using the normalized residual
\[
\Delta z \;=\; \frac{z_{\rm phot}-z_{\rm spec}}{1+z_{\rm spec}},
\]
and report the mean bias $\langle \Delta z \rangle$, the normalized median absolute deviation ($\sigma_{\mathrm{NMAD}}$),
\[
\sigma_{\mathrm{NMAD}} \;=\; 1.4826 \,\mathrm{median}\!\left(\left|\,\Delta z - \mathrm{median}(\Delta z)\,\right|\right),
\]
and the outlier fraction $\eta$, defined as the fraction of objects with $|\Delta z| > 0.15$ \citep{2010A&A...523A..31H}. Fig.~\ref{fig:photoz} (left) compares the photometric redshifts $z_{\rm phot}$ predicted by the conditional diffusion model with the spectroscopic redshifts $z_{\rm spec}$ for the test set, yielding $\sigma_{\rm NMAD}=0.035$ and an outlier fraction of $\eta=7.65\%$.

For comparison, we evaluate our method against CatBoost\footnote{\url{https://github.com/catboost/catboost}}, a gradient-boosted decision-tree method \citep{prokhorenkova2018catboost} that has been successfully applied to photometric redshift estimation \citep{2024AJ....168..233L}. The CatBoost model is trained and evaluated using the same photometric flux inputs and the same training, validation, and test sets described above.
We tune the CatBoost model by grid-searching the tree depth, \texttt{depth} \(\in \{4,8,12,16\}\), and the L2 regularization coefficient, \texttt{l2\_leaf\_reg} \(\in \{1,3,5,7\}\). The final configuration is selected by minimizing \(\sigma_{\rm NMAD}\) on the validation set. During training, we set \texttt{iterations}=50{,}000 and used early stopping with a patience of 100 rounds. We optimized the model with the mean absolute error (MAE) loss and kept other hyperparameters at their default values.
The resulting CatBoost model achieves \(\sigma_{\rm NMAD}=0.037\) and \(\eta=7.18\%\) on the test set (Fig.~\ref{fig:photoz}, right), closely matching the result of the conditional diffusion model. This result indicates that, in terms of point-estimation accuracy, the conditional diffusion model is competitive with a strong tree-based benchmark, while additionally providing per-object uncertainty estimates through posterior sampling.

Fig.~\ref{fig:pit} presents a quantitative assessment of the probabilistic calibration of the photometric-redshift posteriors predicted by the conditional diffusion model on the test set.
The top panel shows the distribution of the probability integral transform (PIT), a commonly used diagnostic for assessing the calibration of probabilistic predictions \citep{gneiting2007probabilistic}, defined as the cumulative distribution function (CDF) of the predicted redshift posterior evaluated at the true spectroscopic redshift. For a well-calibrated model, the PIT values should follow a uniform distribution on $[0,1]$. The bottom panel shows the corresponding probability--probability (P--P) plot, which compares the empirical CDF of the PIT values with the ideal one-to-one relation.
The PIT histogram is approximately uniform, and the empirical P--P curve lies close to the diagonal, indicating that the predicted posteriors are reasonably well calibrated with respect to the spectroscopic redshifts. A mild excess of PIT values near unity suggests a small residual asymmetry, with the true redshifts occasionally falling in the upper tail of the predicted posteriors, consistent with a slight tendency to underestimate $z$ for a subset of objects. Overall, these diagnostics indicate that the sampled posteriors provide a reliable characterization of redshift uncertainties.

\begin{figure}[!t] 
  \centering
  \includegraphics[width=\columnwidth]{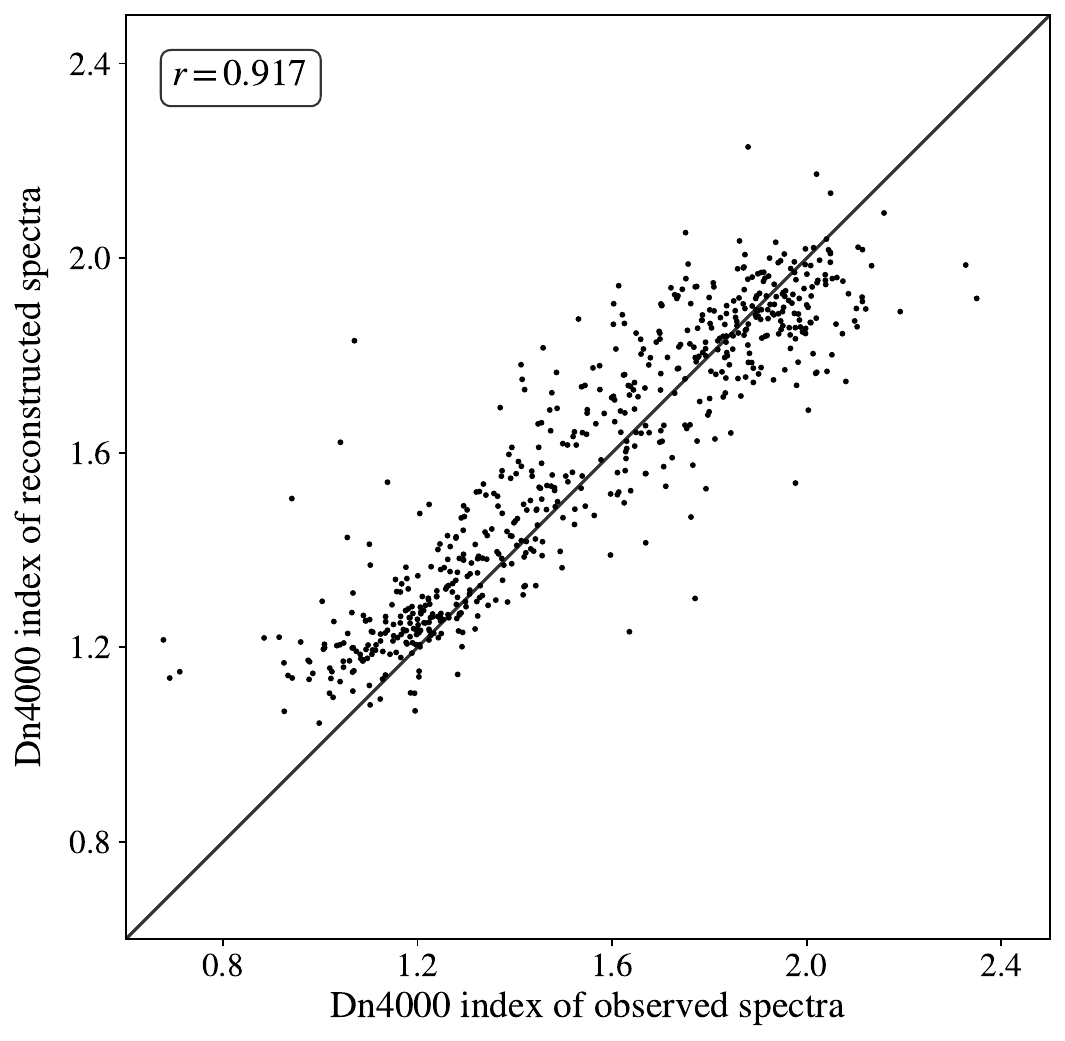}
  \caption{Comparison of the Dn4000 index measured from observed spectra and reconstructed spectra. The solid line indicates the one-to-one relation. The Pearson correlation coefficient is $r = 0.917$.}
  \label{fig:Dn4000}
\end{figure}

\begin{figure*}[!t]
\centering
\includegraphics[width=\textwidth]{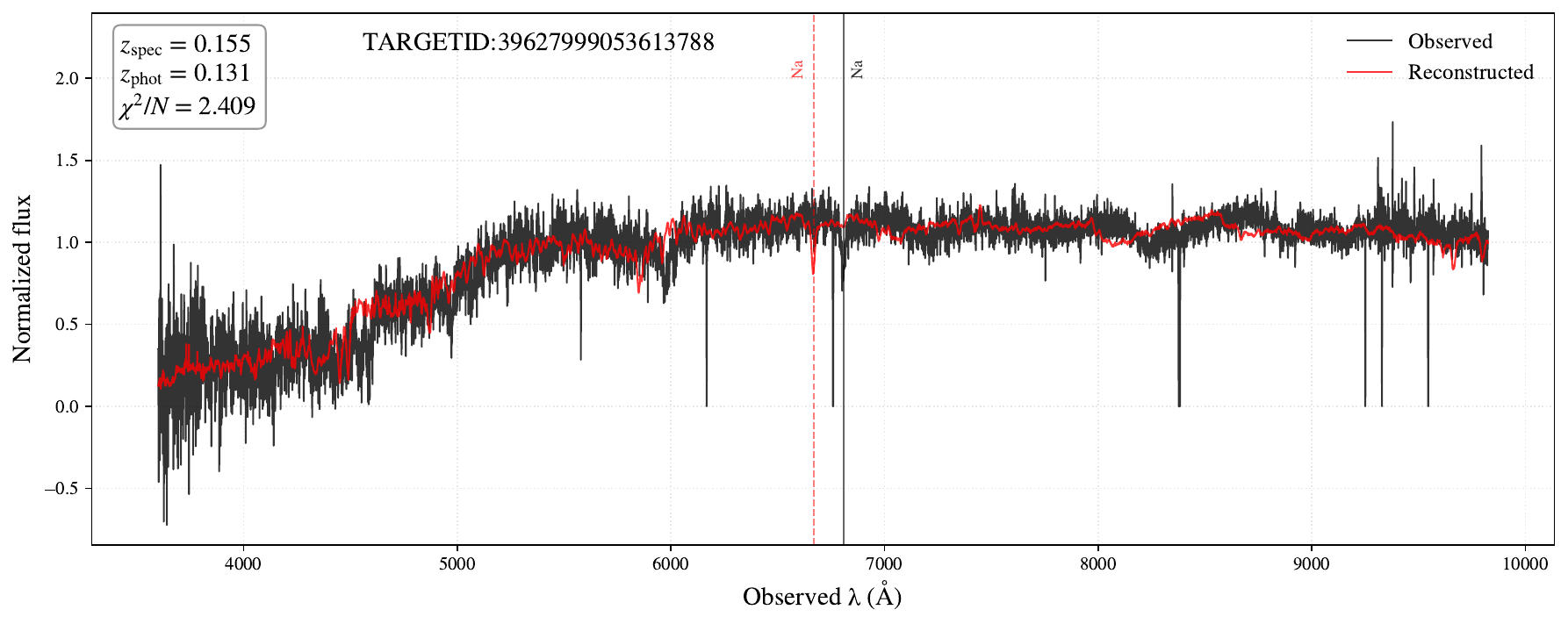}  
\caption{Observed-frame spectrum of an example galaxy. The observed spectrum is shown in black, and the reconstructed spectrum is shown in red. The solid black and dashed red vertical lines mark the expected observed-frame positions of the absorption feature (Na) for \(z_{\rm spec}\) and \(z_{\rm phot}\), respectively. The spectroscopic redshift, photometric redshift, and $\chi^2/N$ are reported in the upper-left corner.}
\label{fig:obsframe_spectrum}
\end{figure*}

\subsection{Reconstructing rest-frame spectra}
\label{sec:rest-frame spectra}

For spectral reconstruction, we decode the spectral latent vectors predicted by the conditional diffusion model using the trained \textsc{spender} decoder to generate rest-frame spectra. To account for uncertainty in the predicted spectra, we draw ten independent samples from the conditional diffusion model. Each sample yields a spectral latent realization, which is then decoded into a rest-frame spectrum. The final reconstructed spectrum is obtained by averaging the flux values of the resulting spectra at each wavelength grid point. This averaged spectrum is used for the spectral reconstruction analysis below.

Fig.~\ref{fig:rest-frame_spectra} presents a visual comparison between the observed spectra (black) and the reconstructed spectra (red) in the rest frame. The first five rows show examples randomly selected from distinct spectroscopic-redshift intervals and ordered by increasing spectroscopic redshift. The bottom row shows an example with a high reconstruction residual.
Each row corresponds to one galaxy: the left panel shows its full spectrum, including the \texttt{TARGETID}, spectroscopic redshift $z_{\rm spec}$ and reconstruction residual, while the three columns to the right provide zoomed-in views of selected wavelength regions with the main emission and absorption features indicated.
For the first galaxy, prominent emission and absorption features that are clearly visible in the observed spectrum are also present in the reconstructed spectrum. The main features appear near the expected wavelengths, although the detailed strengths of individual features are not matched exactly, and the continuum shape is reproduced over most of the wavelength range. For the second galaxy, the reconstructed spectrum also reproduces the overall continuum shape and broadly follows the weaker structures around the Ca II H and K, H$\beta$, and H$\alpha$ regions. The third and fourth examples correspond to noisier spectra at intermediate and higher redshifts, where individual spectral features are less clearly defined. In both cases, the reconstruction mainly follows the broad continuum level and does not introduce strong spurious features. The fifth galaxy is a noisy high-redshift case in which [O II] is the only clearly visible feature. The reconstruction follows the broad continuum level while retaining a feature near the expected [O II] wavelength. Taken together, these examples indicate that the model reproduces the continuum shape over most of the wavelength range and can capture the presence of prominent spectral features.

Given the limited spectral information contained in the five-band photometric data, a perfect reconstruction of individual galaxy spectra is not expected. The sixth row of Fig.~\ref{fig:rest-frame_spectra} shows a high-residual case, randomly selected from the 1\% of objects with the largest \(\chi^2/N\) values (Eq.~\ref{eq:chi2_per_pixel}). 
Its reconstruction shows a noticeable continuum bias and an emission feature near H$\beta$ that is not present in the observed spectrum.
To assess reconstruction fidelity at the population level, we reconstruct the rest-frame spectra for the 10k reconstruction sample and compute $\chi^2/N$ for each object. 
The resulting distribution is shown in the left panel of Fig.~\ref{fig:chiqdist}: the median and mean are \(1.04\) and \(1.23\), respectively, indicating that the reconstruction residuals are generally close to the noise level of the observed spectra.
The distribution also exhibits a high-$\chi^2/N$ tail, but it contains only a small fraction of the sample. The corresponding observed-frame distribution is shown in the right panel of Fig.~\ref{fig:chiqdist} and is discussed further in Sect.~\ref{sec:obs-frame spectra}.

\subsection{Reconstruction of the Dn4000 index}
\label{sec:Dn4000}

To assess whether the reconstructed spectra retain physically meaningful continuum information at the population level, we compare the 4000\,\AA{} break index (Dn4000) measured from reconstructed and observed spectra. Here, the reconstructed spectra refer to the photometry-inferred rest-frame spectra described in Sect.~\ref{sec:rest-frame spectra}.
Dn4000 quantifies the strength of the 4000\,\AA{} continuum break and is widely used as a tracer of stellar population properties \citep{2003MNRAS.341...33K}. Following \citet{1999ApJ...527...54B}, we define the narrow 4000\,\AA{} break index in the rest frame as
\[
{\rm Dn4000}\;=\;\frac{\langle F_\nu\rangle_{4000\text{--}4100\,\text{\AA}}}
{\langle F_\nu\rangle_{3850\text{--}3950\,\text{\AA}}}\,,
\]
where $\langle F_\nu\rangle_{a\text{--}b}$ denotes the mean flux density over the rest-frame wavelength interval $[a,b]$. Since DESI spectra are delivered as $F_\lambda$, we convert the flux density from $F_\lambda$ to $F_\nu$ via $F_\nu(\nu)=\frac{\lambda^2}{c}\,F_\lambda(\lambda)$, with $\lambda=c/\nu$ and $c$ the speed of light.
To ensure robust reference measurements, we restrict the analysis to galaxies whose observed spectra have sufficient signal in the Dn4000 bands, requiring a mean signal-to-noise ratio $\ge 3$ across both wavelength intervals.
Fig.~\ref{fig:Dn4000} compares Dn4000 values measured from the reconstructed and observed spectra from the 10k reconstruction sample. We find a strong Pearson correlation coefficient ($r=0.917$), indicating that the reconstruction preserves the strength of the 4000\,\AA{} break at the population level and that the reconstructed spectra retain physically meaningful continuum information.

\subsection{Reconstruction in the observed frame}
\label{sec:obs-frame spectra}

The rest-frame comparison isolates the quality of the inferred spectral shape, whereas the observed-frame comparison also includes the effect of photometric-redshift uncertainty. Although the full posterior allows multiple spectral realizations through sampling, we adopt a single representative prediction here to enable direct object-by-object comparison in the observed frame. We take the photometric-redshift estimate $z_{\rm phot}$ as the posterior median of the redshift distribution (Sect.~\ref{sec:photoz_prediction}) and use the averaged reconstructed rest-frame spectrum described in Sect.~\ref{sec:rest-frame spectra}. This spectrum is then redshifted to the observed frame using $z_{\rm phot}$, yielding a deterministic prediction for direct comparison with the observed data while reflecting the central tendency of the posterior.

Fig.~\ref{fig:obsframe_spectrum} compares the observed spectrum (black) and the reconstructed spectrum (red) in the observed frame for a representative galaxy. The selected spectrum has a high signal-to-noise ratio and exhibits a well-defined continuum, facilitating a clear comparison.
We mark the absorption feature (Na) to illustrate the effect of redshift uncertainty. 
The solid black vertical line shows its expected observed-frame position for \(z_{\rm spec}=0.155\), while the dashed red vertical line shows the position of the corresponding reconstructed feature when the rest-frame reconstruction is shifted using \(z_{\rm phot}=0.131\).
A visible offset between these markers is evident, arising from the difference between the photometric and spectroscopic redshifts. Despite this shift, the reconstructed spectrum broadly follows the observed continuum shape and reproduces the corresponding absorption feature. 
This indicates that the reconstruction retains key intrinsic spectral information, while the observed-frame alignment of spectral features remains limited by photometric-redshift uncertainty.

We quantify this effect by comparing the $\chi^2/N$ distributions for the rest-frame (Fig.~\ref{fig:chiqdist}, left) and observed-frame (Fig.~\ref{fig:chiqdist}, right) reconstructions.
While the rest-frame reconstruction achieves a median $\chi^2/N$ of $1.04$, the observed-frame results show a slight degradation, with the median increasing to $1.14$ and the mean rising from $1.23$ to $1.56$.
The observed-frame distribution also exhibits a more pronounced high-$\chi^2/N$ tail.
This behavior is expected, as uncertainties in the photometric-redshift estimate introduce small wavelength shifts when the reconstructed rest-frame spectrum is mapped to the observed frame, thereby increasing the reconstruction residual.

\section{Summary and conclusions}\label{sec:summary}

We have presented a generative framework for the joint probabilistic inference of galaxy redshifts and rest-frame spectra from photometric fluxes. This approach recasts photometric redshift estimation from a one-dimensional regression task as a joint generative inference problem over redshift and spectrum. It combines a pre-trained spectral autoencoder, \textsc{spender}, with a conditional diffusion model that maps photometric fluxes to a joint distribution over redshift and a spectral latent representation. The latent representation is subsequently decoded into a rest-frame spectrum using the trained \textsc{spender} decoder. We demonstrate the method on the DESI DR1 data set, a large spectroscopic sample spanning a broad range of galaxy types and redshifts.

For redshift inference, the model provides a photometric-redshift PDF for each galaxy by sampling from the conditional diffusion model, enabling the characterization of non-Gaussian and potentially multimodal uncertainties. The resulting point estimates derived from these posteriors achieve a precision comparable to that of a strong machine-learning baseline (CatBoost) when using the same photometric inputs. PIT diagnostics indicate that the predicted posteriors are well calibrated.

For spectral reconstruction, the model produces rest-frame spectra that can be redshifted and resampled into the observed frame by an explicit transformation. Given the limited information content of broadband photometry, exact reconstruction is not expected. Nevertheless, for most galaxies, the reconstructed spectra reproduce the overall continuum shape and capture the presence of prominent spectral features. Quantitatively, the reconstruction residuals are close to the observational noise level, with a mean chi-square per pixel $\chi^2/N$ of $1.23$ on a reconstruction sample. For galaxies with sufficient signal-to-noise, the Dn4000 index shows good agreement between reconstructed and observed spectra, indicating that key continuum properties are preserved at the population level.

We emphasize several limitations of this work. First, broadband photometry provides only coarse sampling of galaxy SED information, and the inferred results therefore remain affected by color–redshift and color–SED degeneracies. 
In the present study, these limitations are influenced by the adopted photometric coverage (three optical and two mid-infrared bands), with the achievable spectral and redshift constraints depending directly on the filter set and wavelength coverage.
Future applications to surveys with broader and deeper photometric information, such as \textit{Euclid}, LSST, or combined multi-survey datasets, offer a path to reduce these degeneracies and improve both spectral and redshift inference.

Second, the reconstructed spectra are constrained to a learned manifold defined by the autoencoder. While this regularization enables stable inference from limited photometric information, it also implies that the reconstructions reflect assumptions encoded in the training data and the learned representation. The inferred spectra should therefore be interpreted as probabilistic realizations consistent with both the photometry and the learned spectral prior. Although the reconstructions capture population-level spectral trends, their use for precise inference of detailed galaxy physical properties (e.g., stellar populations or emission-line diagnostics) requires further validation. Despite these limitations, the proposed framework provides a promising basis for extracting spectral information from large photometric surveys in a statistically consistent manner.

\begin{acknowledgements}

We thank Jose Antonio Alcolea López for sharing \textsc{spender} training scripts and for helpful discussions that informed this work. Han-Yue Guo is supported by the program of China Scholarship Council (Grant No.202306030074). Martin Eriksen acknowledges support from the Spanish Ministry of Science and Innovation through the project PID2023-152069NA-I00. This work has received funding from the ‘Severo Ochoa Centres of Excellence’ programme (grant CEX2024-001441-S), funded by MICIU/AEI/10.13039/501100011033.
This research used data obtained with the Dark Energy Spectroscopic Instrument (DESI). DESI construction and operations is managed by the Lawrence Berkeley National Laboratory. This material is based upon work supported by the U.S. Department of Energy, Office of Science, Office of High-Energy Physics, under Contract No. DE–AC02–05CH11231, and by the National Energy Research Scientific Computing Center, a DOE Office of Science User Facility under the same contract. Additional support for DESI was provided by the U.S. National Science Foundation (NSF), Division of Astronomical Sciences under Contract No. AST-0950945 to the NSF’s National Optical-Infrared Astronomy Research Laboratory; the Science and Technology Facilities Council of the United Kingdom; the Gordon and Betty Moore Foundation; the Heising-Simons Foundation; the French Alternative Energies and Atomic Energy Commission (CEA); the National Council of Humanities, Science and Technology of Mexico (CONAHCYT); the Ministry of Science and Innovation of Spain (MICINN), and by the DESI Member Institutions: \url{www.desi.lbl.gov/collaborating-institutions}. The DESI collaboration is honored to be permitted to conduct scientific research on I’oligam Du’ag (Kitt Peak), a mountain with particular significance to the Tohono O’odham Nation. Any opinions, findings, and conclusions or recommendations expressed in this material are those of the author(s) and do not necessarily reflect the views of the U.S. National Science Foundation, the U.S. Department of Energy, or any of the listed funding agencies.
The DESI Legacy Imaging Surveys consist of three individual and complementary projects: the Dark Energy Camera Legacy Survey (DECaLS), the Beijing-Arizona Sky Survey (BASS), and the Mayall z-band Legacy Survey (MzLS). DECaLS, BASS and MzLS together include data obtained, respectively, at the Blanco telescope, Cerro Tololo Inter-American Observatory, NSF’s NOIRLab; the Bok telescope, Steward Observatory, University of Arizona; and the Mayall telescope, Kitt Peak National Observatory, NOIRLab. NOIRLab is operated by the Association of Universities for Research in Astronomy (AURA) under a cooperative agreement with the National Science Foundation. Pipeline processing and analyses of the data were supported by NOIRLab and the Lawrence Berkeley National Laboratory (LBNL). Legacy Surveys also uses data products from the Near-Earth Object Wide-field Infrared Survey Explorer (NEOWISE), a project of the Jet Propulsion Laboratory/California Institute of Technology, funded by the National Aeronautics and Space Administration. Legacy Surveys was supported by: the Director, Office of Science, Office of High Energy Physics of the U.S. Department of Energy; the National Energy Research Scientific Computing Center, a DOE Office of Science User Facility; the U.S. National Science Foundation, Division of Astronomical Sciences; the National Astronomical Observatories of China, the Chinese Academy of Sciences and the Chinese National Natural Science Foundation. LBNL is managed by the Regents of the University of California under contract to the U.S. Department of Energy. The complete acknowledgments can be found at \url{https://www.legacysurvey.org/acknowledgment/}.

\end{acknowledgements}

%

\bibliographystyle{bibtex/aa} 
\bibliography{refs}







   
  



\end{document}